\documentclass{article}

\usepackage{amssymb}
\usepackage{amsmath}
\usepackage{epsfig}

\begin{document}

\title{\bf High-Speed Cylindrical Collapse of Two Perfect Fluids}

\author{M. Sharif \thanks{msharif@math.pu.edu.pk} and Zahid Ahmad
\thanks{zahid$\_$rp@yahoo.com}\\
\textit{\ {\small Department of Mathematics, }} \\
\textit{{\small University of the Punjab, Lahore 54590,
Pakistan}}}
\date{}
\maketitle

\begin{abstract}
In this paper, the study of the gravitational collapse of
cylindrically distributed two perfect fluid system has been
carried out. It is assumed that the collapsing speeds of the two
fluids are very large. We explore this condition by using the
high-speed approximation scheme. There arise two cases, i.e.,
bounded and vanishing of the ratios of the pressures with
densities of two fluids given by $c_s,~d_s$. It is shown that the
high-speed approximation scheme breaks down by non-zero pressures
$p_1,~p_2$ when $c_s,~d_s$ are bounded below by some positive
constants. The failure of the high-speed approximation scheme at
some particular time of the gravitational collapse suggests the
uncertainity on the evolution at and after this time. In the
bounded case, the naked singularity formation seems to be
impossible for the cylindrical two perfect fluids. For the
vanishing case, if a linear equation of state is used, the
high-speed collapse does not break down by the effects of the
pressures and consequently a naked singularity forms.  This work
provides the generalisation of the results already given by Nakao
and Morisawa [1] for the perfect fluid.
\end{abstract}


\section{Introduction}

One of the central issues in the General theory of Relativity has
always been the question of gravitational radiation. Historically,
the first gravitational radiation problem considered was the
radiation from a spinning rod [2,3]. One of the important issues
of gravitational radiation is the gravitational collapse. General
Relativity  suggests that gravitational collapse of massive
objects results in the formation of spacetime singularities in our
universe [4-6]. The known physical laws break down very near to
the spacetime singularity. At this stage, quantum theory of
gravity is applied to describe the physical phenomena in the
neighbourhood of the spacetime singularity. One of the important
issues "whether the spacetime singularities, (far from the region
where the gravitational collapse occurs) formed in our universe,
can be seen by observer or not" has attracted many people.

Penrose [7] investigated this problem and suggested a cosmic
censorship hypothesis. This hypothesis includes two versions, one
is called {\it strong version} and other is called {\it weak
version}. According to the strong version, a spacetime singularity
formed by non-singular initial data is not visible from infinity
while according to the weak version, a spacetime singularity
formed by non-singular initial data is invisible for any observer
[8]. The singularity claimed by the strong version is named as
naked singularity, while the singularity claimed by the weak
version is known as globally naked singularity. Nakamura et al.
[9] conjectured that large spacetime curvatures move away to
infinity in the form of gravitational radiation in the
neighbourhood of globally naked singularities (if exist).

The spherically symmetric systems are considered to be very useful
[10] to discuss the problem of naked singularity formation. These
systems are simple and have well-defined physical significance. We
shall take non-spherical perturbations to this system [11] to
study the problem of gravitational radiation. We consider
cylindrically symmetric system because there is a degree of
gravitational radiation and the spacetime singularity formed in
this system is naked [12,13]. There is a great literature [14-19]
available on gravitational radiation through cylindrical
gravitational collapse. Some numerical work on the gravitational
radiation has also been done by different people [20-22].

Recently, Nakao and Morisawa [1] have discussed the gravitational
collapse of a cylindrical thick shell composed of a perfect fluid
to the case of non-vanishing pressure. Same authors [23] have also
specialized this work for the complete dust case. These
investigations have provided interesting results about the
gravitational collapse. This paper has been addressed for the
high-speed cylindrical collapse of the two perfect fluids. We use
high-speed approximation scheme by considering perturbation
analysis. For this purpose, the collapsing speeds of the two
perfect fluids are assumed almost equal to the speed of light.
Further, we take the deviation of the 4-velocities of the two
fluids from null as a small perturbation. We investigate the
effects of the pressures for the large collapsing velocities. It
is verified that our results reduce to the perfect fluid case as
obtained by Nakao and Morisawa [1].

The rest of the paper is outlined as follows. In section 2, we
give the basic structure of the spacetime with whole-cylinder
symmetry [24]. Section 3 is devoted to discuss the high-speed
approximation scheme for the two perfect fluids. In section 4, we
interpret the effects of pressures of the two fluids on the
high-speed gravitational collapse. Finally, section 5 contains
summary and discussion.

\section{Cylindrically Symmetric Two Perfect Fluid System}

The spacetime with the whole-cylinder symmetry is given by the
line element [24]
\begin{equation}
ds^2=e^{2(\gamma-\psi)}(-dt^2+dr^2)+e^{2\psi}dz^2+e^{-2\psi}R^2d\phi^2,
\end{equation}
where $\gamma$, $\psi$ and $R$ are functions of $t$ and $r$ only.
Einstein field equations (EFEs) are given by
\begin{equation}
R_{ab}-\frac{1}{2}Rg_{ab}=\kappa T_{ab},\quad (a,b=0,1,2,3),
\end{equation}
where $R_{ab}$ is the Ricci tensor, $R$ is the Ricci scalar,
$T_{ab}$ is the energy-momentum tensor and $\kappa$ is the
gravitational constant. These equations for (2.1) take the
following form
\begin{eqnarray}
\gamma'&=&(R'^2-{\dot{R}}^2)^{-1}\{RR'({\dot \psi}^2+{\psi'}^2)
-2R\dot{R}\dot{\psi}\psi'+R'R''-\dot{R}{\dot{R}}'\nonumber\\
&-&\kappa\sqrt{-g}(R'{T^t}_t+\dot{R}{T^r}_t)\},\\
\dot{\gamma}&=&-({R'}^2-{\dot{R}}^2)^{-1}\{R\dot{R}(\dot{\psi}^2+{\psi'}^2)
-2RR'\dot{\psi}\psi'+\dot{R}R''-R'{\dot{R}}'\nonumber\\
&-&\kappa\sqrt{-g}(\dot{R}{T^t}_t+R'{T^r}_t)\},\\
\ddot{\gamma}&-&{\gamma}''={\psi'}^2-{\dot{\psi}}^2
-\frac{\kappa}{R}\sqrt{-g}{T^\phi}_\phi,\\
\ddot{R}&-&R''=-\kappa\sqrt{-g}({T^t}_t+{T^r}_r),\\
\ddot{\psi}&+&\frac{\dot{R}}{R}\dot{\psi}-{\psi}''-\frac{R'}{R}\psi'
=-\frac{\kappa}{2R}\sqrt{-g}({T^t}_t+{T^r}_r-{T^z}_z+{T^\phi}_\phi),
\end{eqnarray}
where dot and prime represent differentiation w.r.t. $t$ and $r$
respectively.

There is a literature available [25,26] which shows interesting
investigation of the combination of two different fluids. e.g. Hall
[25] presented eight different combinations and investigated their
physical as well as mathematical properties. Here we have taken one
of these combination which is more simpler and has obvious
interesting properties due to perfect fluid combination. This system
of two perfect fluid is also used to obtain kinematic self-similar
solutions of the Einstein field equations [27]. For the sake of
simplicity, we have taken only two fluids instead of taking
arbitrary.

We consider the non-interacting combination of two (non-zero)
perfect fluids where their (unit, time-like, future-pointing) flow
vectors $u$ and $v$ are assumed non-parallel. If these fluids have
densities $\rho_1,~\rho_2$ and pressures $p_1,~p_2$, the
energy-momentum tensor for two perfect fluid system [25] is given by
\begin{equation}
T_{ab}=(\rho_{1}+p_{1})u_a u_b+(\rho_{2}+p_{2})v_a
v_b+(p_{1}+p_{2})g_{ab}.
\end{equation}
It is assumed not only that the separate dominant energy
conditions hold, but that the extra conditions $p_j\geq0$ hold and
so $p_j+\rho_j>0,~(j=1,~2)$. The Segre type is $\{1,1(11)\}$ with
corresponding eigenvalues
$-(\rho_1+\rho_2)-\epsilon,~(p_1+p_2)+\epsilon,~p_1+p_2$
(repeated), where
\begin{eqnarray}
2\epsilon&=&-(p_1+p_2+\rho_1+\rho_2)^2+\{(p_1+p_2+\rho_1+\rho_2)^2\nonumber\\
&-&4(p_1+\rho_1)(p_2+\rho_2)[1-(u_a v^a)^2]\}^\frac{1}{2}
\end{eqnarray}
and so $\epsilon>0$. It is obvious that if energy-momentum tensor
represents a perfect fluid with $p+\rho>0$, then it uniquely
determines the fluid flow $u^a$ and the quantities $p$ and $\rho$.
For the combination of two perfect fluids, the physical
characteristics of the fluid represented by the flow vector, the
pressure, and the density are uniquely determined. Since $T_{ab}$
uniquely determines its eigenvalues, $p_1+p_2$ and $\rho_1+\rho_2$
are determined and $p_1+\rho_1$ and $p_2+\rho_2$ are determined to
within an interchange of their values. Consequently, for each
choice of the values of $p_1+\rho_1$ and $p_2+\rho_2$, there is a
one-parameter family of values for the quadruple
$(p_1,~p_2,~\rho_1,~\rho_2)$ restricted by the requirements that
$0\leq p_j\leq\rho_j$ and $\rho_j>0$.

The components of the 4-velocity $u^a$ for the fluid 1 of the
whole-cylinder symmetry take the form [1]
\begin{equation}
u^a=u^t(1,-1+U,0,0).
\end{equation}
Since $u^a$ is time-like, U is positive. Using the normalization
property, i.e., $u^au_a=-1$, $u^t$ is written as
\begin{equation}
u^t=\frac{e^{-\gamma+\psi}}{\sqrt {U(2-U)}}.
\end{equation}
Similarly, for the fluid 2, the 4-velocity will become
\begin{equation}
v^a=v^t(1,-1+V,0,0),\quad (v^av_a=-1)
\end{equation}
where $v^t$ turns out to be
\begin{equation}
v^t=\frac{e^{-\gamma+\psi}}{\sqrt{V(2-V)}} .
\end{equation}
We define the new variables $D_{1}$, $P_{1}$ and $D_{2}$, $P_{2}$
corresponding to the two fluids as
\begin{eqnarray}
D_{1}:&=&\frac{{\sqrt{-g}}(\rho_{1}+p_{1})u^t}{\sqrt{U(2-U)}}=\frac{R
e^{\gamma-\psi}(\rho_{1}+p_{1})}{U(2-U)},\\
D_{2}:&=&\frac{{\sqrt{-g}}(\rho_{2}+p_{2})v^t}{\sqrt{V(2-V)}}=\frac{R
e^{\gamma-\psi}(\rho_{2}+p_{2})}{V(2-V)},\\
P_{1}:&=&\frac{R e^{\gamma-\psi}p_{1}}{U(2-U)},\\
P_{2}:&=&\frac{R e^{\gamma-\psi}p_{2}}{V(2-V)},
\end{eqnarray}
where $g$ represents the determinant of the metric tensor
$g_{ab}$. The non-vanishing components of the energy-momentum
tensor ${T^a}_b$ turn out to be
\begin{eqnarray}
{\sqrt{-g}}{T^t}_t&=&e^{\gamma-\psi}\{-D_{1}-D_{2}+U(2-U)P_{1}+
V(2-V)P_{2}\},\\
{\sqrt{-g}}{T^r}_t&=&e^{\gamma-\psi}\{D_{1}(1-U)+D_{2}(1-V)\}=
-{\sqrt{-g}}{T^t}_r,\\
{\sqrt{-g}}{T^r}_r&=&e^{\gamma-\psi}\{D_{1}(1-U)^2+D_{2}(1-V)^2+
U(2-U)P_{1}\nonumber\\
&+&V(2-V)P_{2}\},\\
{\sqrt{-g}}{T^z}_z&=&e^{\gamma-\psi}\{U(2-U)P_{1}+V(2-V)P_{2}\}=
{\sqrt{-g}}{T^\phi}_\phi.
\end{eqnarray}
Equation for the conservation of energy-momentum, i.e.,
${T^a_b};_a=0$ corresponding to the two fluids can be written in
the component form as
\begin{eqnarray}
\partial_u(D_{1}+D_{2})&=&-\frac{1}{2}(D_{1}U+D_{2}V)'+
\frac{1}{2}\{U(2-U)P_1+V(2-V)P_2 \}^.\nonumber\\
&+&\frac{D_1(1-U)}{2}\{2\partial_u(\psi-\gamma)
-U(\dot{\psi}-\dot{\gamma})\}\nonumber\\
&+&\frac{D_{2}(1-V)}{2}\{2\partial_u(\psi-\gamma)
-V(\dot{\psi}-\dot{\gamma})\}\nonumber\\
&+& \frac{1}{2} \{U(2-U)P_{1}+V(2-V)P_2\}
(\dot{\psi}-\dot{\gamma}-\frac{\dot{R}}{R}),
\end{eqnarray}
\begin{eqnarray}
D_1\partial_uU+D_2\partial_uV&=&(1-U)\partial_uD_1
+(1-V)\partial_uD_2\nonumber\\
&+&\frac{1}{2}[\{U(1-U)D_1+V(1-V)D_2\}\nonumber\\
&-&\{U(2-U)P_1+V(2-V)P_2\}]'\nonumber\\
&-&\frac{D_1}{2}\{2\partial_u(\psi-\gamma)
-U(\dot{\psi}-\dot{\gamma})\}\nonumber\\
&-&\frac{D_2}{2}\{2\partial_u(\psi-\gamma)
-V(\dot{\psi}-\dot{\gamma})\}\nonumber\\
&+&\frac{1}{2}\{U(2-U)P_1+V(2-V)P_2\}(\gamma'-\psi'+\frac{R'}{R}),
\end{eqnarray}
where $u=t-r$ is the retarded time, $v=t+r$ is the advanced time
and $\partial_u$ is the partial derivative w.r.t. $u$. The first
of these equations is obtained from the $t$-component and the
second is obtained from the $r$-component. The $z$- and
$\phi$-components vanish.

The energy flux of gravitational radiation can be found by
considering the C-energy E and its flux vector. The C-energy is
defined as [24]
\begin{equation}
E=\frac{1}{8}\{1+e^{-2\gamma}(\dot{R}^2-{R'}^2)\},
\end{equation}
where $E=E(t,r)$.The corresponding energy flux vector $J^a$ is defined as
\begin{equation}
\sqrt{-g}J^a=\frac{4}{\kappa}(E',-\dot{E},0,0).
\end{equation}
where $J^a$ satisfies the conversation law. When we make use of
Eqs.(2.3), (2.4) and (2.24) in Eq.(2.25), the C-energy flux vector
can be written in component form as
\begin{eqnarray}
\sqrt{-g}J^t=\frac{e^{-2\gamma}}{\kappa}\{RR'({\dot
\psi}^2+{\psi'}^2)-2R\dot{R}\dot{\psi}\psi'
-\kappa\sqrt{-g}(R'{T^t}_t+\dot{R}{T^r}_t)\},\\
\sqrt{-g}J^r=\frac{e^{-2\gamma}}{\kappa}\{R\dot{R}(\dot{\psi}^2
+{\psi'}^2)-2RR'\dot{\psi}\psi'
-\kappa\sqrt{-g}(R'{T^r}_t-\dot{R}{T^r}_r)\}.
\end{eqnarray}
The remaining components vanish.

\section{High-Speed Approximation Scheme}

In this section we use linear perturbation analysis w.r.t. a small
parameter $\epsilon$. This is called high-speed approximation
scheme. Since we are considering two perfect fluid system, we take
the ingoing null limits of the cylindrical two perfect fluids.
Using $D_{1},~D_{2}$ and $P_{1},~P_{2}$, the energy-momentum
tensor can be written in the form
\begin{equation}{\setcounter{equation}{1}}
T_{ab}=\frac{e^{3(\psi-\gamma)}}{R}[D_1 k_a k_b+D_2 l_a
l_b+e^{2(\gamma-\psi)}\{U(2-U)P_1+V(2-V)P_2\}g_{ab}],
\end{equation}
where
\begin{eqnarray}
k^a&=&(1,-1+U,0,0),\\
l^a&=&(1,-1+V,0,0).
\end{eqnarray}
In the limits $U\rightarrow 0_+,~V\rightarrow 0_+$, the time-like
vectors $k^a$ and $l^a$ become null vectors. Keeping
$D_1,~D_2,~P_1,~P_2$ finite with these limits, the energy-momentum
tensor coincides with the collapsing two null dusts, i.e.,
\begin{equation}
T_{ab}\rightarrow\frac{e^{3(\psi-\gamma)}}{R}[D_1 k_a k_b+D_2 l_a
l_b],
\end{equation}
where
\begin{eqnarray}
k^a&\rightarrow&(1,-1,0,0),\\
l^a&\rightarrow&(1,-1,0,0).
\end{eqnarray}
\begin{figure}
\center \epsfig{file=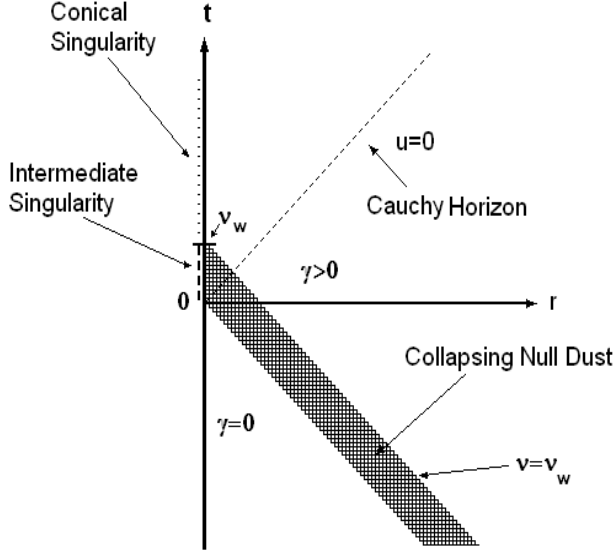,width=0.695\linewidth} \caption{The
figure shows Morgan's cylindrical null dust solution. The null dust
is shown in the shaded region, where $D(v)>0$ for $0<v<v_W$. The
dashed line on t-axis corresponds to the intermediate singularity at
which an observer suffers infinite tidal force although any scalar
polynomials of the Riemann tensor do not vanish. The dotted line on
t-axis is the conical singularity. The short dashed line $t=r$ is
the Cauchy horizon.}
\end{figure}

It follows that the two perfect fluid system is approximated by a
two null dust system for very large collapsing velocities, i.e.,
$0<U\ll 1,~0<V\ll 1$. We take the deviation of the four-velocities
from a null  as a perturbation and perform linear perturbation
analysis.

It can be found that for the collapsing two null dusts, the
solution is given by
\begin{eqnarray}
\psi&=&0,\\
\gamma&=&\gamma_B(v),\\
R&=&r,\\
\kappa(D_1+D_2)e^\gamma&=&\frac{d\gamma_B}{dv},
\end{eqnarray}
where $\gamma_B$ is an arbitrary function of $v$. This solution has
been investigated by many people [14,16,17]. We take this solution
as a background spacetime for the perturbation analysis. It follows
from Eqs .(3.4) and (3.9) that the energy-momentum tensor diverges
at the symmetry axis $r=0$ if $D_1$ and $D_2$ do not vanish
simultaneously. Using Einstein field equations, the same behaviour
can also be shown for the Ricci tensor. This gives rise to a naked
singularity. It is mentioned here that the same results hold for the
perfect fluid case [1] with either $D_1=0$ or $D_2=0$. Some previous
studies [16] show that although the Kretschmann scalar vanishes, the
tidal forces felt by freely falling test particles may become
unbounded. This indicates that a non-scalar or intermediate
singularity is formed there [28,29]. The singularity structure can
be shown in the figure 1.

Now we take large collapsing velocities of the two perfect fluids
and use linear perturbation analysis. We assume that orders of the
variables are such that $U=O(\epsilon),~V=O(\epsilon)$ and
$\psi=O(\epsilon)$. The variables $\gamma,~R$ and $D_1,~D_2$ can
be written as
\begin{eqnarray}
e^\gamma&=&e^{\gamma_B}(1+\delta_\gamma),\\
R&=&r(1+\delta_R),\\
D_1&=&D_B(1+\delta_{D_1}),\\
D_2&=&D_B(1+\delta_{D_2}),
\end{eqnarray}
where $\delta_\gamma,~\delta_R$ and $\delta_{D_1},~\delta_{D_2}$
are of $O(\epsilon)$ and $D_B$ is given by
\begin{equation}
D_B:=\frac{1}{2\kappa e^{\gamma_B}}\frac{d\gamma_B}{dv}.
\end{equation}
It is mentioned here that $B$ stands for background spacetime in
the perturbation analysis and $\gamma_B(v)$ is an arbitrary
function of the advanced time $v$. We expand the variables
$U,~V,~\psi,~\delta_\gamma,~\delta_R,~\delta_{D_1},~\delta_{D_2}$
up to first order w.r.t. $\epsilon$. Eqs.(2.3)-(2.7) and (2.22)
take the following forms respectively.
\begin{eqnarray}
{\delta_\gamma}'&=&2\kappa D_Be^{\gamma_B}\{\delta_\gamma-\psi+
\frac{(\delta_{D_1}+\delta_{D_2})}{2}\nonumber\\
&-&\frac{(UP_1+VP_2)}{D_B}-2\partial_v(r\delta_R)\}+(r\delta_R)'',
\end{eqnarray}
\begin{eqnarray}
\dot{\delta_\gamma}&=&2\kappa D_Be^{\gamma_B}\{\delta_\gamma-\psi
+\frac{(\delta_{D_1}+\delta_{D_2})}{2}\nonumber\\
&-&\frac{(U+V)}{2}-2\partial_v(r\delta_R)\}+(r\dot{\delta_R})',\\
\ddot{\delta_\gamma}&-&{\delta_\gamma}''=-\frac{2\kappa e^{\gamma_B}}{r}(UP_1+VP_2),\\
r\ddot{\delta_R}&-&(r\delta_R)''=2\kappa e^{\gamma_B}\{(D_B-2P_1)U+(D_B-2P_2)V\},\\
\ddot{\psi}&-&\psi''-\frac{1}{r}\psi'=\frac{\kappa
e^{\gamma_B}}{r}\{(D_B-2P_1)U+(D_B-2P_2)V\},
\end{eqnarray}
\begin{eqnarray}
\partial_u(\delta_{D_1}+\delta_{D_2}+2\delta_\gamma-2\psi)
&=&-\frac{1}{2}\{(U+V)'\
-\frac{d\gamma_B}{dv}(U+V)\}\nonumber\\
&+&\frac{1}{D_B}\{(UP_1+VP_2)^.
-(UP_1+VP_2)\frac{d\gamma_B}{dv}\}\nonumber\\
&-&\frac{1}{2D_B}\frac{dD_B}{dv}(U+V).
\end{eqnarray}
Using Eq.(3.21), Eq.(2.23) takes the form
\begin{equation}
\partial_u\{(D_B-2P_1)U+(D_B-2P_2)V\}=\frac{UP_1+VP_2}{r}.
\end{equation}
The first-order expression for the C-energy w.r.t. $\epsilon$
becomes
\begin{equation}
E=\frac{1}{8}[1-e^{-2\gamma_B}+2e^{-2\gamma_B}\{\delta_\gamma-(r\delta_R)'\}].
\end{equation}
From Eq.(3,15), it is clear that $\gamma_B$ is constant in the
region where $D_B=0$, i.e., in the vacuum region. Then it follows
from Eqs.(3.16) and (3.17) that $\delta_\gamma-(r\delta_R)'$ is
constant in the vacuum region. Thus up to the first order, the
C-energy is constant in the vacuum region. This implies that, up
to the first order in $\epsilon$, the C-energy flux vector $J^a$
vanishes and up to the second-order in $\epsilon$, it is given in
component form as
\begin{eqnarray}
\sqrt{-g}J^t&=&\frac{r}{\kappa}({\dot{\psi}}^2+{\psi'}^2),\\
\sqrt{-g}J^r&=&-\frac{2r}{\kappa}\dot{\psi}\psi'.
\end{eqnarray}
It follows from here that the C-energy flux vector corresponds to
the massless Klein-Gordon field.

\section{Effects of Pressure on the High-Speed Collapse}

In this section, we discuss the consequences of the pressures on
the high-speed collapse. We take energy densities $\rho_1,~\rho_2$
and pressures $p_1,~p_2$ of the two fluids positive and define the
following quantities
\begin{eqnarray}{\setcounter{equation}{1}}
c_s(u,v)&=&\sqrt{\frac{p_1}{\rho_1}},\\
d_s(u,v)&=&\sqrt{\frac{p_2}{\rho_2}},
\end{eqnarray}
where the subscript $s$ stands for speed. When the dominant energy
conditions are satisfied, $c_s$ and $d_s$ are less than or equal to
the speed of light [30]. In this paper, we assume that $c_s$ and
$d_s$ are such that
\begin{eqnarray}
c_s<1,\\
d_s<1.
\end{eqnarray}
It will be shown below that the high-speed approximation does not
work  for the cases of $c_s=1$ and $d_s=1$. From Eqs.(2.14),
(2.16) and (4.1), it follows that
\begin{equation}
P_1=\frac{{c_s}^2}{1+{c_s}^2}D_1.
\end{equation}
Also, Eqs.(2.15), (2.17) and (4.2) yield
\begin{equation}
P_2=\frac{{d_s}^2}{1+{d_s}^2}D_2.
\end{equation}
Using Eqs.(4.5) and (4.6) in Eq.(3.22), we get
\begin{equation}
\partial_u\{(\frac{1-{c_s}^2}{1+{c_s}^2})U
+(\frac{1-{d_s}^2}{1+{d_s}^2})V\}
=\frac{U{c_s}^2}{(1+{c_s}^2)r}+\frac{V{d_s}^2}{(1+{d_s}^2)r}.
\end{equation}
Integrating this equation, it turns out
\begin{eqnarray}
(\frac{1-{c_s}^2}{1+{c_s}^2})U+(\frac{1-{d_s}^2}{1+{d_s}^2})V
&=&\int^{u}_{\tilde{U}(v)}\frac{2U{c_s}^2(x,v)}
{\{1+{c_s}^2(x,v)\}(v-x)}dx\nonumber\\
&+&\int^{u}_{\tilde{U}(v)}\frac{2V{d_s}^2(x,v)}
{\{1+{d_s}^2(x,v)\}(v-x)}dx,
\end{eqnarray}
where $\tilde{U}(v)$ is an arbitrary function of the advanced time
$v$. Eq.(4.8) can be re-written in the following two forms either
\begin{eqnarray}
U+(\frac{1+{c_s}^2}{1-{c_s}^2})(\frac{1-{d_s}^2}{1+{d_s}^2})V
&=&(\frac{1+{c_s}^2}{1-{c_s}^2})[\int^{u}_{\tilde{U}(v)}\frac{2U{c_s}^2(x,v)}
{\{1+{c_s}^2(x,v)\}(v-x)}dx\nonumber\\
&+&\int^{u}_{\tilde{U}(v)}\frac{2V{d_s}^2(x,v)}{\{1+{d_s}^2(x,v)\}(v-x)}dx],
\end{eqnarray}
or
\begin{eqnarray}
(\frac{1+{d_s}^2}{1-{d_s}^2})(\frac{1-{c_s}^2}{1+{c_s}^2})U+V
&=&(\frac{1+{d_s}^2}{1-{d_s}^2})[\int^{u}_{\tilde{U}(v)}\frac{2U{c_s}^2(x,v)}
{\{1+{c_s}^2(x,v)\}(v-x)}dx\nonumber\\&+&\int^{u}_{\tilde{U}(v)}
\frac{2V{d_s}^2(x,v)} {\{1+{d_s}^2(x,v)\}(v-x)}dx].
\end{eqnarray}
It can be seen from Eqs.(4.9) and (4.10) that the velocity
perturbations $U,~V$ are not defined for $c_s=1$ and $d_s=1$
respectively. Thus the high-speed approximation scheme ceases to
hold for these values of $c_s,~d_s$.

Now we consider the following two cases: \\
(i) When $c_s$ and $d_s$
are bounded below by some positive
constant; \\
(ii) When $c_s$ and $d_s$ vanish for
$\rho_1\rightarrow\infty,~\rho_2\rightarrow\infty$.

\subsection{Bounded Case}

Here we take $c_s$ and $d_s$ as bounded below by some positive
constants $e$ and $f$ respectively. It follows that the following
inequalities hold
\begin{eqnarray}
\frac{2{c_s}^2}{1+{c_s}^2}\geq{e^2},\\
\frac{2{d_s}^2}{1+{d_s}^2}\geq{f^2},
\end{eqnarray}
where $e$ and $f$ are non-vanishing constants. Using Eqs.(4.11)
and (4.12) in Eq.(4.7), we obtain
\begin{equation}
\partial_u\{(\frac{1-{c_s}^2}{1+{c_s}^2})U+(\frac{1-{d_s}^2}{1+{d_s}^2})V \}\geq
\frac{Ue^2}{v-u}+\frac{Vf^2}{v-u}.
\end{equation}
Integrating this inequality, we have
\begin{eqnarray}
(\frac{1-{c_s}^2}{1+{c_s}^2})U&+&(\frac{1-{d_s}^2}{1+{d_s}^2})V\geq
ln\{const\times{r}^{-(Ue^2+Vf^2)}\}\nonumber\\
&+&e^2\int\frac{dU}{du}\ln(v-u)du+f^2\int\frac{dV}{du}\ln(v-u)du,
\end{eqnarray}
where the $constant$ is positive. This shows that the perturbation
velocities $U,~V$ diverge at the symmetric axis $r=0$.
Consequently, the high-speed approximation scheme breaks down in
this case.

\subsection{Vanishing Case}

In this case, we investigate the behavior of the two perfect
fluids when $c_s,~d_s$ vanish for the high energy density limits,
i.e., $\rho_1\rightarrow\infty,~\rho_2\rightarrow\infty$. The
energy densities will become larger and larger when the fluid
elements approach the background singularity, i.e., $r=0$, (the
region of $D_B\neq0$). Thus the asymptotic behaviors of $c_s$ and
$d_s$ turn out to be
\begin{eqnarray}
{c_s}^2(u,v)\sim{C_s}^2(t)r^q={C_s}^2(v-r)r^q\sim{C_s}^2(v)r^q,\\
{d_s}^2(u,v)\sim{D_s}^2(t)r^q={D_s}^2(v-r)r^q\sim{D_s}^2(v)r^q,
\end{eqnarray}
where $C_s,~D_s$ are arbitrary functions and $q$ is positive
constant. It is mentioned here that $D_s$ is different from
$D_1,~D_2$ defined in section $2$. Using the above asymptotic
forms of $c_s$ and $d_s$ in Eq.(4.8), we get
\begin{eqnarray}
\{\frac{1-{C_s}^2(v)r^q}{1+{C_s}^2(v)r^q}\}U+\{\frac{1-{D_s}^2(v)r^q}
{1+{D_s}^2(v)r^q}\}V\sim &-&\frac{2{C_s}^2(v)}{q}[U(u,v)r^q
-U\{\tilde{U}(v),v\}
(\frac{v-\tilde{U}}{2})^q]\nonumber\\
&-&\frac{2{D_s}^2(v)}{q}[V(u,v)r^q-V\{\tilde{U}(v),v\}
(\frac{v-\tilde{U}}{2})^q]\nonumber\\
&+&\frac{2}{q}[\int^{u}_{\tilde{U}(v)}\frac{dU}{dx}(x,v)
\ln\{1+{C_s}^2(v)(\frac{v-x}{2})^q\}dx\nonumber\\
&+&\int^{u}_{\tilde{U}(v)}\frac{dV}{dx}(x,v)\ln\{1
+{D_s}^2(v)(\frac{v-x}{2})^q\}dx].
\end{eqnarray}
From here we see that the perturbation velocities $U,~V$ remain
finite at the background singularity, i.e., $r=0$. Consequently,
the high-speed approximation scheme works well in this case. This
shows that the naked singularity forms at the background
singularity through this cylindrical gravitational collapse.

Now we consider the asymptotic equations of state which have the
same asymptotic behavior as Eqs.(4.15) and (4.16). From Eqs.(2.14)
and (2.15), it follows that the energy densities $\rho_1,~\rho_2$
take the following forms
\begin{eqnarray}
\rho_1\sim\frac{2D_BU}{re^{\gamma_B}},\\
\rho_2\sim\frac{2D_BV}{re^{\gamma_B}}
\end{eqnarray}
at the background singularity. Adding Eqs.(4.18) and (4.19) and
use Eq.(4.17) at the background singularity, we obtain
\begin{equation}
\rho_1+\rho_2\sim\frac{2D_B}{re^{\gamma_B}}(U+V)=(function~of
~v)^2\times\frac{1}{r}.
\end{equation}
When we make use of Eqs.(4.1), (4.2), (4.15), (4.16) in the above
equation, it follows that
\begin{equation}
p_1+p_2={c_s}^2\rho_1+{d_s}^2\rho_2\sim
const\times(\rho_1+\rho_2)^{1-q}
\end{equation}
where $constant$ is positive. This yields a linear equation of
state. Thus the high-speed collapse does not break by the effects
of pressures if linear equation of state is used. This forms a
spacetime naked singularity.

\subsection{Generation of Gravitational Radiation}

It can be seen from Eq.(3.20) that the right hand side represents
the source term. This source term contains perturbation velocities
$U,~V$ and if these velocities become larger, $|\psi|$ also
becomes larger. Due to these large perturbations of $U$ and $V$,
the C-energy flux in the the asymptotically flat region, given by
Eqs.(3.24) and (3.25), implies that large amount of gravitational
radiations emit.

The bounded $c_s,~d_s$ case indicates that the perturbation
velocities $U,~V$ become unbounded at the background singularity.
For the high-speed approximation scheme, the perturbation
velocities are meaningful only less than unity. This shows that a
significant amount of gravitational radiation is emitted by the
pressure deceleration at the background singularity. Our result
agrees with Piran's numerical results [20] which indicate that the
large amount of gravitational radiation is emitted when the
pressure bounce occurs.

It has been found that for vanishing $c_s,~d_s$ case that the
perturbation velocities $U,~V$ remain finite even at the
background singularity. This result is in agreement with the dust
fluid case for the generation of gravitational radiation [23].

\section{Outlook}

Gravitational collapse is one of the most striking phenomena in
gravitational physics. The cosmic censorship has provided strong
motivation for researchers in this field. It is interesting to
investigate the gravitational waves from a naked singularity
formed by the high-speed gravitational collapse by using different
symmetries.

This paper is devoted to investigate the gravitational collapse of
the cylindrically symmetric two perfect fluids using the
high-speed approximation scheme. We have considered two cases,
i.e., bounded and vanishing of $c_s,~d_s$. It is shown that the
high-speed approximation scheme breaks down  by non-zero pressures
$p_1,~p_2$ when $c_s,~d_s$ are bounded below by some positive
constants. The failure of the high-speed approximation scheme at
some particular time of the gravitational collapse suggests the
uncertainity on the evolution at and after this time. In the
bounded case, the naked singularity formation seems to be
impossible for the cylindrical two perfect fluids. For the
vanishing case, it is shown that in the limit of high energy
densities, $\rho_1\rightarrow\infty,~\rho_2\rightarrow\infty$, the
high-speed collapse continues until a naked singularity forms.
Alternatively, if a linear equation of state is used such that
$p_1+p_2$ is proportional to $(\rho_1+\rho_2)^{1-q}$ with $0<q<1$,
the high-speed collapse does not break down by the effects of
pressures and consequently a naked singularity forms.

It follows from Eq.(3.20) that the large amount of gravitational
radiation can be measured with the increase of the velocity
perturbations  $U$ and $V$. This means that the cylindrical two
perfect fluids with $c_s$ and $d_s$ bounded below by some positive
values will emit more gravitational waves in its collapse due to the
unbounded increase in $U$ and $V$ which breaks down the high speed
approximation scheme. However, for the vanishing of $c_s,~d_s$ in
the high energy density limit, $U,~V$ are finite even at the
background singularity. The vanishing of $c_s,~d_s$ would generate
the larger amount of gravitational radiation keeping the
perturbation velocities finite. Also, high speed approximation
scheme remains valid here. This shows the similarity of the behavior
with the dust case and consequently generates the larger amount of
gravitational radiation. It is mentioned here that if we take
$\rho_1=0=p_1$ or $\rho_2=0=p_2$ all the results reduce to the
cylindrical perfect fluid case [1]. It would be interesting to
explore this problem by taking some other symmetry.

\newpage

\vspace{0.5cm}

{\bf Acknowledgments}

\vspace{0.5cm}

 We acknowledge the enabling role of the Higher
Education Commission Islamabad, Pakistan, and appreciate its
financial support through the {\it Indigenous PhD 5000 Fellowship
Program Batch-I}. We are thankful to Miss Umber Sheikh for helping
us in sketching the diagram. We are also thankful for the
anonymous referees for their useful comments.

\end{document}